\documentclass{elsart}
\usepackage{graphicx}

\begin{document}

\begin{frontmatter}

\title{Microscopic analysis of $T=1$ and $T=0$ proton-neutron correlations
       in $N=Z$ nuclei}

\author[Sawara]{M. Hasegawa} and
\author[Sandai]{K. Kaneko} 

\address[Sawara]{Laboratory of Physics, Fukuoka Dental College,
 Fukuoka 814-0193, Japan}
\address[Sandai]{Department of Physics, Kyushu Sangyo University,
 Fukuoka 813-8503, Japan}

\begin{abstract}

  The competition between the isovector ($T=1$) and isoscalar
($T=0$) proton-neutron ($p-n$) correlations in $N=Z$ nuclei
is investigated by calculating their correlation energies
with a realistic effective interaction which reproduces observed nuclear
properties very well, in a strict shell model treatment.
It is shown  in the realistic shell model that the double-differences
of binding energies 
$(B(A+pn:T) + B(A) - B(A+p) - B(A+n))$ ($B(A)$ being the binding energy)
are good indicators of the $T=1$ and $T=0$ $p-n$ correlations.
Each of them, however, originates in plural kinds of correlations
with $T=1$ or $T=0$.

\vspace*{3mm}
\leftline{PACS: 21.10.-k,21.10.Dr,21.60.Cs}

\end{abstract}

\end{frontmatter}

\section{Introduction}

   The competition between isovector ($T=1$) and isoscalar ($T=0$) pairing
correlations has been a matter of renewed concern in nuclear structure
studies of $N \approx Z$ nuclei \cite{Zeldes,Macc,Satula1}.
The competition appears in the near degeneracy of the lowest $T=1$ and $T=0$
states in odd-odd $N=Z$ nuclei.  
The $T=0$ proton-neutron ($p-n$) pairing correlations in $N \approx Z$ nuclei
have been studied by the two approaches \cite{Macc,Satula1}
with different treatments of the symmetry energy. 
The two conclusions about the importance of the $T=0$ $p-n$ pairing
correlations are in opposition to one another.
The $T=1$ and $T=0$ pairing correlations, which are considered to be induced
by $T=1$ and $T=0$ nuclear interactions, should be treated consistently
on the same footing \cite{Satula1}.
The structure of $N=Z$ nuclei has been well described by the shell model
which does treat $T=1$ and $T=0$ pairing consistently.
  Large-scale shell model calculations were applied to the investigation
 of the isovector and isoscalar pairing correlations in Refs.
 \cite{Poves,Mart,Sat},
 where the contributions of  $T=1$, $J=0$  and $T=0$, $J=1$ interactions
 are compared \cite{Poves} and the contributions of the quadrupole-quadrupole
 ($QQ$) force are also considered \cite{Mart}.
 The authors have recently shown that the competition between the $T=1$
 and $T=0$ pairing correlations are approximately explained with the $T=1$,
 $J=0$ pairing force ($P_0$) and a $T=0$ $p-n$ force ($V^{T=0}_{mp}$ below)
 \cite{Kaneko4}.
 In order to understand the competition in detail, it is important
 to evaluate the two types of correlations induced by realistic effective
 interactions.

   In this paper, we investigate the competition between $T=1$ and $T=0$
$p-n$ correlations in the lowest states of $N=Z$ nuclei
using a realistic effective interaction in a strict shell model treatment.
 The spherical shell model, which gives an excellent description of
various properties of $N \approx Z$ nuclei (not only the binding
energies but also other properties),
has the advantage that the correlation energies of respective interactions
are properly calculated.
   A shell model Hamiltonian is composed of the $T=0$ and
 $T=1$ interactions,
\begin{eqnarray}
 & {} & H = H_{sp} + V^{T=0} + V^{T=1}, \\  \label{eq:1}
 & {} & V^{T} = \sum_{a \leq b} \sum_{c \leq d} \sum_{JM} \sum_{TK}
     G_{JT}(ab:cd) A^\dagger_{JMTK}(ab) A_{JMTK}(cd)
   \quad (T=0, 1), \label{eq:2}
\end{eqnarray}
where $H_{sp}$ stands for the single-particle energies,
$A^\dagger_{JMTK}(ab)$ is the creation operator of a nucleon pair
with the spin $JM$ and the isospin $TK$ in the single-particle orbits
($a,b$), and $G_{JT}(ab:cd)$ denotes the interaction matrix elements.
  The so-called realistic effective interactions contain multipole
($J \ge 0$) pairing forces of $T=0$ and $T=1$ in the expression (\ref{eq:2}).
In this sense, the shell model with a realistic effective interaction is
to deal with all the {\it multipole pairing correlations}.
 We investigate the competition between the $T=0$ and $T=1$ correlations
induced by the $T=0$ and $T=1$ interactions.

   The realistic effective interactions have the property
that the centroid of $T=0$ diagonal interaction matrix elements 
$\overline{G_{T=0}(ab)}=\sum_J (2J+1)G_{J0}(ab:ab)$ $/\sum_J (2J+1)$ has
a roughly constant value, being independent on the orbits $(ab)$ \cite{Kaneko3}.
By setting $-k_0 = \sum_{ab} \overline{G_{T=0}(ab)} / \sum_{ab}$,
we obtain the average $T=0$ $p-n$ force
\begin{equation}
 V^{T=0}_{mp} = - k^0 \sum_{a \leq b} \sum_{JM}
         A^\dagger_{JM00}(ab) A_{JM00}(ab).         \label{eq:3}
\end{equation}
Let us write residual $T=0$ interactions as
$V^{T=0}_{res} = V^{T=0} - V^{T=0}_{mp}$ and rewrite 
the Hamiltonian as
\begin{equation}
  H = H_{sp} + V^{T=0}_{mp} + V^{T=0}_{res} + V^{T=1}. 
        \label{eq:4}
\end{equation}
 The separation of $V^{T=0}_{mp}$ in Eq. (\ref{eq:4}) follows
the procedure of Dufour and Zuker in Ref. \cite{Zuker},
where the Hamiltonian is divided into the monopole and multipole parts
as $H=H_m+H_M$.
 The monopole field $V^{T=0}_{mp}$ is exactly written as
$-(k^0 / 2) \{ {\hat n_v}/2 ({\hat n_v}/2 +1) - {\hat T}({\hat T}+1) \}$,
where ${\hat n_v}$ stands for the number of valence nucleons and
${\hat T}$ stands for the total isospin.
  This equation shows that the force $V^{T=0}_{mp}$ which yields
only energy gain depending on $n_v$ and $T$ is completely irrelevant
to wave-functions, even to whether the nucleus is spherical or deformed,
and configuration mixing is caused by $V^{T=0}_{res}$ and $V^{T=1}$.
The force strength $k^0$ extracted from the realistic effective interaction
USD \cite{Wilden} is, for instance, 2.8 MeV for $^{22}$Na.
We know that $V^{T=0}_{mp}$ makes a large contribution to
the binding energy and gives a bonus 2.8 MeV to the $T=0$ states
against the $T=1$ states in $^{22}$Na.
In this paper, we do not separate the $T=1$ monopole field
which is much smaller than the $T=0$ one
(we shall mention another reason for it later).
We analyze roles of the three interactions $V^{T=0}_{mp}$,
$V^{T=0}_{res}$ and $V^{T=1}$ in the competition between
the $T=0$ and $T=1$ correlations.

\section{Indicators of $T=0$ and $T=1$ $p-n$ correlations}

   It is well known that the even-even $N=Z$ nuclei are more deeply bound
than the $T=1$, $0^+$ states of neighboring nuclei.  The difference in mass
between the nuclei with mass number $A=m \alpha$ and $A=m \alpha +2$
(where the unit $\alpha$ consists of two protons ($2p$) and two neutrons ($2n$)
and $m$ is an integer)
 is a good indicator of the $\alpha$-like $2p-2n$ correlations according
  to Ref. \cite{Gambhir}. 
If the difference is large, it shows the $\alpha$-like superfluidity
of the $A=m \alpha$ system.  In fact, we see such a signature
in the whole region of $N \approx Z$ nuclei.
 We can regard the $T=1$ (or $T=0$) lowest state of the odd-odd $N=Z$ nuclei
 as a correlated state of the last $p-n$ pair coupled with the $\alpha$-like
 superfluid state of $A=m \alpha$.
 Based on this picture, we attempted to analyze the single-difference
 of binding energies $B(m \alpha +pn)-B(m \alpha)$,
 but we were unable to obtain simple information about the $T=1$ or $T=0$
 correlations of the last $p-n$ pair.

   The $p-n$ correlation energy in the odd-odd $N=Z$ nuclei
with $A=m \alpha+pn$ can be evaluated by the double-difference of
binding energies \cite{Janecke,Jensen,Kaneko2}
\begin{eqnarray}
 -\delta ^2V_{pn} (A + pn:T) = (B(A+pn:T) + B(A)) \nonumber \\ 
  \hspace{1cm}          - (B(A+p) + B(A+n)) \quad (T=0,1),   \label{eq:5}
\end{eqnarray}
where $B(A)$ stands for the binding energy.
It is popular to calculate the odd-even mass difference for evaluation
of the $T=1$ pairing correlations. Using the odd-even mass difference
$\Delta M_{oe}(A+n)$, the correlation energy of the $T=1$ neutron pair
in the $A+2n$ (even) nuclei is evaluated by
\begin{eqnarray}
 & { } & -\delta ^2V_{nn}(A+2n) = -2 \Delta M_{oe}(A+n)  \nonumber \\ 
 & { } & \hspace{1.5cm}  =  B(A+2n) + B(A) - 2B(A+n).   \label{eq:6}
\end{eqnarray}
In the BCS approximation for the $T=1$, $J=0$ pairing correlations,
the odd-$A$ system is described as one neutron quasi-particle
on the pairing superfluid vacuum as the average state of adjacent
even-$A$ systems, and the quasi-particle energy $E_{qp}$ is
approximately equal to the gap $\Delta _{n}$,
{\it i.e.}, $\Delta M_{oe} \approx E_{qp} \approx \Delta _{n}$.
Since the isospin is a good quantum number in $N \approx Z$ nuclei,
the $m\alpha +2p$, $m\alpha +2n$ and $m\alpha +pn$ nuclei with $T=1$
(or $m \alpha +p$ and $m \alpha +n$ nuclei) have approximately
the same energy if the Coulomb energy is subtracted from the binding energy.
 We have the approximate relation
$ -\delta ^2V_{pn}(m \alpha + pn:T=1) \approx
   -\delta ^2V_{nn}(m \alpha + 2n)$.
We can suppose that the quantity $-\delta ^2V_{pn}(m \alpha + pn:T=1)$
represents the correlation energy of the last $T=1$ $p-n$ pair
as $-\delta ^2V_{nn}(m \alpha + 2n)$ does that of the last neutron pair.
Similarly, $-\delta ^2V_{pn}(m \alpha + pn:T=0)$ represents
the correlation energy of the last $p-n$ pair with $T=0$ outside the
$A=m \alpha$ system.

\begin{figure}[b]
\begin{center}
\includegraphics[width=6.5cm,height=6.5cm]{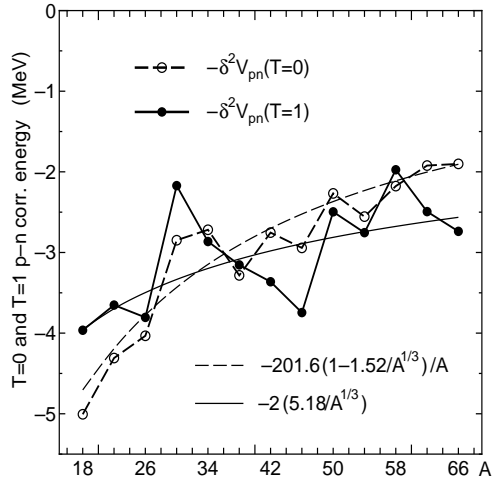}
  \caption{Empirical values of $ -\delta ^2V_{pn}$ for the lowest
           $T=0$ and $T=1$ states of odd-odd $N=Z$ nuclei.}
  \label{fig1}
\end{center}
\end{figure}

   Let us illustrate empirical values of
$-\delta ^2V_{pn}(T=0)$ and $-\delta ^2V_{pn}(T=1)$ for the lowest
$T=0$ and $T=1$ states of odd-odd $N=Z$ nuclei, in Fig. \ref{fig1}.
The difference $\delta ^2V_{pn}(T=0) - \delta ^2V_{pn}(T=1)$ is equal to
the energy difference between the lowest $T=0$ and $T=1$ states, from the
definition (\ref{eq:5}).  It is considered in phenomenology
that the quantities $-\delta ^2V_{pn}(T=0)$ and $-\delta ^2V_{pn}(T=1)$
are indicators of the $T=0$ and $T=1$ pairing correlations.
The odd-even mass difference expressed as $5.18A^{-1/3}$
in the mass formula \cite{Duflo} corresponds well with
$\delta ^2V_{pn}(T=1)/2$, as shown in Fig. \ref{fig1}.
If we write the symmetry energy plus Wigner energy as $a_T T(T+1)$
in a mass formula, we have the relation
$\delta ^2V_{pn}(T=0) \approx 3a_T/2$ (note that if the Wigner energy
proportional to $T$ is neglected, $\delta ^2V_{pn}(T=0) \approx a_T/2$).
The empirical value $a_T \approx 134.4 (1 - 1.52 A^{-1/3})$ in the mass
formula \cite{Duflo} explains the $A$-dependent value of
$\delta ^2V_{pn}(T=0)$ in Fig. \ref{fig1}.
  The empirical values of $\delta ^2V_{pn}(T=1)$ and
  $\delta ^2V_{pn}(T=0)$ show large fluctuations, which indicates
  shell effects. Basically, however, $\delta ^2V_{pn}(T=0)$ is
  larger than $\delta ^2V_{pn}(T=1)$ in the $sd$ shell nuclei,
  while the latter is larger than the former in the
  $pf$ shell nuclei. The empirical trend depending on the mass $A$ is
  well explained by using the parameters of the mass formula \cite{Duflo}.
  The trend of $\delta ^2V_{pn}(T=0)$ is also reproduced by $3 a_T/2$ 
  with somewhat different parameters of other modern mass formulas
  which have the symmetry energy and Wigner energy in the form
  $a_T T(T+1)$.
 This suggests an intimate relation between the $T=0$ correlations
and the symmetry energy. 
 The model calculations using the $T=1$ pairing force
and a $T=0$ $p-n$ force on a deformed base \cite{Satula1} and on a spherical
base \cite{Kaneko4} gave a microscopic explanation of the competition
between the $T=1$ and $T=0$ correlations, reproducing the energy difference 
$\delta ^2V_{pn}(T=0) - \delta ^2V_{pn}(T=1)$.

\begin{figure}[b]
\begin{center}
\includegraphics[width=7cm,height=7cm]{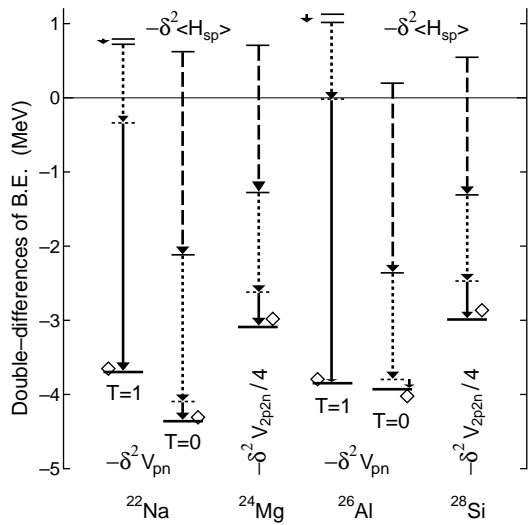}
  \caption{The double-differences of binding energies
           $ -\delta ^2V_{pn}(m \alpha +pn:T)$ and
           $ -\delta ^2V_{2p2n}(m \alpha +2p2n)$.
           The starting point of $ -\delta ^2V_{pn}$
           or $ -\delta ^2V_{2p2n}$ with no interactions
           is shown by the top line
           ($ -\delta^2 \langle H_{sp} \rangle $) in each column.
           The respective contributions from $V^{T=0}_{mp}$,
           $V^{T=0}_{res}$ and $V^{T=1}$ are denoted by
           the dash, dot and solid lines.
           The open diamonds denote the empirical values.}
  \label{fig2}
\end{center}
\end{figure}

   Our subject is to analyze in more detail the $p-n$
correlation energies $-\delta ^2V_{pn}(T=0)$ and $-\delta ^2V_{pn}(T=1)$
in the shell model calculations with realistic effective interactions.
Let us illustrate the results obtained with the USD interaction
\cite{Wilden} for the $sd$-shell nuclei, in Fig. \ref{fig2},
where contributions from $V^{T=0}_{mp}$, $V^{T=0}_{res}$ and $V^{T=1}$
to $-\delta ^2V_{pn}$ are denoted by the dash, dot and solid lines
respectively, and $- \delta ^2 \langle H_{sp} \rangle$ denotes
the contribution from $H_{sp}$.
 The double-difference of single-particle energies
 $- \delta ^2 \langle H_{sp} \rangle$ is evaluated by substituting
 the expectation value $\langle H_{sp} \rangle$ for $B$ in Eq. (\ref{eq:5}).
 The quantity $- \delta ^2 \langle H_{sp} \rangle$ gives the starting point
 of $-\delta ^2V_{pn}$ with no interactions in the shell model calculation.
In Fig. \ref{fig2}, the calculated values of $-\delta ^2V_{pn}(T=0)$ and
$-\delta ^2V_{pn}(T=1)$ finely reproduce their empirical values.
We see mixed contributions from the $T=0$ and $T=1$
interactions to the final states with $T=0$ or $T=1$ and some deviations
coming from the single-particle energy part $H_{sp}$. 
 Still, Fig. \ref{fig2} shows that the double-differences of binding
energies $-\delta ^2V_{pn}(T=0)$ and $-\delta ^2V_{pn}(T=1)$
are good indicators of the $T=0$ and $T=1$ $p-n$  correlations.
It is notable that not only $V^{T=0}_{mp}$ but also $V_{res}^{T=0}$
contributes to $-\delta ^2V_{pn}(T=0)$. 
We calculated separately the contributions from the $T=1$, $J=0$ interactions
and the other ($T=1$, $J>0$) interactions to the $T=1$ $p-n$ correlation
 energy. The calculated indicator $-\delta ^2V_{pn}(T=1)$
includes the contributions from the $T=1$, $J>0$ interactions as well as
the $T=1$, $J=0$ interactions. 
The results show that the $T=1$, $J>0$ interactions contribute greatly
to the binding energy but reduce the value of $-\delta ^2V_{pn}(T=1)$
considerably.
If we consider only the $T=1$, $J=0$ pairing force $P_0$
and adjust the force strength so as to explain the quantities
$\delta ^2V_{pn}(T=1)$ and $\delta ^2V_{nn}$ for the last nucleon pair, 
the binding energy will not be reproduced, revealing a contradiction.

   We can evaluate the $p-n$ correlation energy in even-even $N=Z$ nuclei
using the following double-difference of binding energies
\cite{Kaneko3,Kaneko2,Zhang,Brenner,Zaochun}:
\begin{eqnarray}
 -\delta ^2V_{2p2n}(A + 2p2n) =  (B(A+2p2n) + B(A)) \nonumber \\ 
 \hspace{1.5cm}       - (B(A+2p) + B(A+2n)).      \label{eq:7}
\end{eqnarray}
 When we consider the $A=m \alpha$ systems,
the quantity $\delta ^2V_{2p2n}$ is nothing but the difference in mass
caused by the $\alpha$-like correlations \cite{Gambhir,Hase3}.
The large values of $\delta ^2V_{2p2n}$ observed in the even-even $N=Z$
nuclei show the $\alpha$-like superfluidity.  The definition of
$\delta ^2V_{2p2n}$ based on the picture of the $\alpha$-like superfluidity
corresponds well with that of $\delta ^2V_{nn}$ based on the pairing
superfluidity.
The quantity $-\delta ^2V_{2p2n}$ represents
the $p-n$ correlation energy between $2p$ and $2n$. 
A quarter of $-\delta ^2V_{2p2n}$ ($-\delta ^2V_{2p2n}/4$) is regarded
as the $p-n$ correlation energy per one $p-n$ bond.

   In Fig. \ref{fig2}, we show the calculated and empirical values of
$-\delta ^2V_{2p2n}/4$ for the $sd$-shell nuclei $^{24}$Mg and $^{28}$Si.
This figure indicates that the quantity $-\delta ^2V_{2p2n}/4$ evaluates
the $T=0$ $p-n$ correlation energy as the quantity $-\delta ^2V_{pn}(T=0)$.
The value of $\delta ^2V_{2p2n}/4$ is smaller than that of $\delta ^2V_{pn}(T=0)$,
which means that the $p-n$ correlations of the last $p-n$ pair with $T=0$
in the $A=m \alpha +pn$ nuclei are a little stronger than the $p-n$
correlations between $2p$ and $2n$. 
 By the way, $\delta ^2V_{2p2n}$ corresponds to the symmetry energy
of the $A+2p$ or $A+2n$ system with $T=1$ in the framework
of the mass formula \cite{Leander}.
Thus, the indicators $\delta ^2V_{pn}(T=0)$ and $\delta ^2V_{2p2n}$
represent the $T=0$ $p-n$ correlation energy in the shell model, 
while they are attributed to the symmetry energy in the mass formulas.
If one excludes the symmetry energy from the binding energy as done in
Ref. \cite{Macc}, one inevitably misses the contribution from the $T=0$
$p-n$ interactions to the total energy and hence is likely to miss
the signature of the $T=0$ correlations.

\section{Further discussions of $T=0$ and $T=1$ correlations}

\begin{figure}[b]
\begin{center}
\includegraphics[width=7cm,height=7cm]{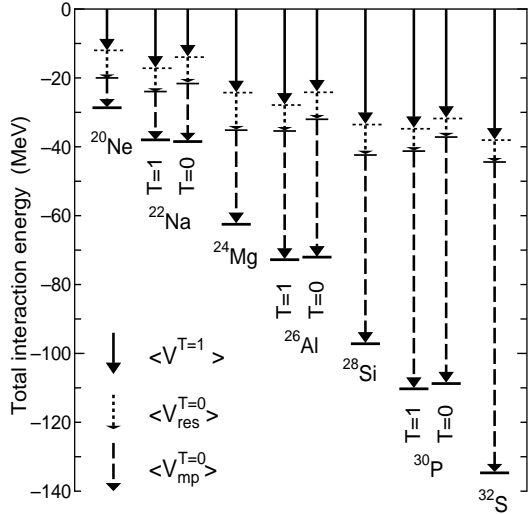}
  \caption{Interaction energies of $V^{T=0}_{mp}$, $V^{T=0}_{res}$
           and $V^{T=1}$ in the $sd$-shell nuclei.
           Their sum is the total interaction energy.}
  \label{fig3}
\end{center}
\end{figure}

It is instructive for our study of the $T=1$ and $T=0$ correlations
to see the expectation values of $V^{T=0}_{mp}$, $V^{T=0}_{res}$
and $V^{T=1}$ in addition to the double-differences of binding energies.
We calculated their expectation values in the lowest states with $T=0$
and $T=1$.  We call the expectation values ``interaction energies"
and write them such as $\langle V^{T=1} \rangle$.
The interaction energies obtained by using the USD interaction 
for the $N=Z$ $sd$-shell nuclei from $^{20}$Ne to $^{32}$S are shown
with three different lines in Fig. \ref{fig3}. 
 This figure shows the indivisible cooperation of the $T=1$
 correlations and the $T=0$ $p-n$ correlations,
and indicates a significant role of the average $T=0$ $p-n$ force
$V^{T=0}_{mp}$ in the realistic effective interaction. 
The larger the valence-nucleon number is, the greater is
the $V^{T=0}_{mp}$ contribution to the binding energy.
   Figure 3 shows that the $T=0$ $p-n$ interactions including 
$V^{T=0}_{mp}$ make a larger contribution to the binding energy
than the $T=1$ interactions.
However, if we neglect $V^{T=0}_{mp}$ which does not affect the structure,
the $T=1$ interactions are more important than the $T=0$ residual
interactions.

   Dufour and Zuker \cite{Zuker} indicated that the realistic effective
interactions are represented approximately by the monopole field $V_{mp}$, 
the $P_0$ force and the $QQ$ force.  In fact, it was shown
in Ref. \cite{Hase1} that the interaction $V_{mp}+P_0+QQ$
with the $T=1$, $J=2$ pairing force ($P_2$) reproduces various properties
of the $f_{7/2}$-subshell nuclei as comparably as the realistic effective
interactions. The extended $P+QQ$ interaction describes well
also the $sd$-shell nuclei in our calculations (unpublished).
 These results showed that the essential part of the monopole field is
 the $T=0$ one ($V_{mp}^{T=0}$) and the $T=1$ monopole field
 can be neglected, although some monopole corrections ($\Delta V_{mp}^{T=0}$
 and $\Delta V_{mp}^{T=1}$) depending on respective orbits remain.
 This is another reason why we do not extract the $T=1$ monopole field
 in Eq. (\ref{eq:4}).
The realistic effective interactions are well approximated by
\begin{eqnarray}
 V_{mp}^{T=0} + \Delta V_{mp}^{T=0}
             + \Delta V_{mp}^{T=1} + P_0 +P_2 +QQ . \label{eq:8}
\end{eqnarray}
 We have a useful insight into the $T=0$ and $T=1$ correlations in this model.
The $T=0$ correlations are induced by $T=0$ re-coupling terms of the $QQ$
force and $\Delta V_{mp}^{T=0}$ except for $V_{mp}^{T=0}$,
 while the $T=1$ correlations are induced by the monopole
and quadrupole pairing forces ($P_0$ and $P_2$), and also by $T=1$ re-coupling
terms of the $QQ$ force and $\Delta V_{mp}^{T=1}$.

   As is well-known, the $T=0$ states are the ground states in the odd-odd
$N=Z$ nuclei except for $^{34}$Cl in the $sd$ shell.  It is curious
that the interaction energy gain of the $T=1$, $0^+$ state is
larger than that of the lowest $T=0$ state for $^{26}$Al and $^{30}$P
in Fig. \ref{fig3}.
 In the shell model, the $T=0$ states of $^{26}$Al and $^{30}$P suffer
less energy loss due to the single-particle energy part
 $\langle H_{sp} \rangle$ and hence become the ground states.
This happens because the $T=1$ correlations are stronger than the $T=0$
residual $p-n$ ones, thus making more nucleons jump up to the $d_{3/2}$ orbit. 
If we lower the $d_{3/2}$ orbit, the $T=1$ state becomes lowest
in $^{26}$Al.  The shell structure thus affects the competition between
the $T=0$ and $T=1$ states for the ground-state position in the odd-odd
$N=Z$ nuclei.

\section{Conclusion}

   In conclusion, we have investigated the competition
of the $T=0$ and $T=1$ $p-n$ correlations in an exact treatment
of the shell model with a realistic effective interaction.
By dividing the $T=0$ $p-n$ interactions into the average monopole field
and the residual interactions, we evaluated interaction energies of them
and the $T=1$ interactions.
The calculations show that the cooperation of the average $T=0$ $p-n$ force,
$T=0$ residual $p-n$ interactions and $T=1$ interactions
produce the large binding energies of the $N=Z$ nuclei.
It is shown that the double-difference of binding energies
$-\delta ^2V_{pn}(T=0)$ or $-\delta ^2V_{pn}(T=1)$
defined by Eq. (\ref{eq:5}) can be used to evaluate the correlation
energy of the last $p-n$ pair with $T=0$ or $T=1$ in the odd-odd $N=Z$ nuclei.
The realistic shell model calculations, however, show that the indicator
$-\delta ^2V_{pn}(T=0)$ or $-\delta ^2V_{pn}(T=1)$ does not originate
in a single kind of pairing correlations, but contains the contributions
from plural kinds of correlations with $T=0$ or $T=1$.
  
  This work was stimulated by the workshop on ``isoscalar and isovector
 pairing" organized by Prof. R. Wyss in Stockholm in May 2003.
 The authors are grateful to Dr. D. M. Brink for his help in improving
 the manuscript.




\end{document}